\documentclass[12pt]{article}
\usepackage{graphicx,epsfig}
\usepackage{color}
\usepackage{cite}
\title{Pressure induced valence and metal-insulator transitions 
in the Falicov-Kimball model with nonlocal hybridization}
\author{Pavol Farka\v sovsk\'y\\
Institute  of  Experimental  Physics,  Slovak   Academy   of
Sciences\\
Watsonova 47, 043 53 Ko\v {s}ice, Slovakia}
\date{}
\begin{document}
\baselineskip=24pt
\maketitle

\begin{abstract}
We present a simple, but very realistic, model for a description 
of pressure induced valence and metal-insulator transitions
in mixed valence systems. It is based on the extended Falicov-Kimball 
model and the supposition that the key interaction governing these transitions 
is the nonlocal hybridization between the localized $f$ and itinerant 
$d$ electrons. Taking into account, in addition,  the~parametrization 
between the external pressure and the $d$-$f$ hybridization 
(the experimental fact), the model is able to describe, at least 
qualitatively, valence as well as  metal-insulator transitions driven by 
the external pressure observed experimentally in some rare-earth
systems, like SmB$_6$.
\end{abstract}

\newpage
\section{Introduction}
The valence and metal-insulator transitions belong certainly
to the most popular manifestations of cooperative phenomena
in solids. These transitions are observed in a wide group
of substances formed by transition-metal oxides as well as 
rare-earth sulphides, halides and borides, when some external
parameters (like pressure or temperature) are varied~\cite{Cho}.
They are in many cases first-order phase transitions, however,
second-order transitions ranging from very gradual to rather
steep are also observed.

To describe all such  transitions  in a unified picture
Falicov and Kimball~\cite{Fal} introduced a  simple  model in
which  only  two  relevant single-electron states  are  taken
into  account:  extended Bloch waves and a set of localized  
states centered  at  the sites of the metallic ions in the crystal.
It is assumed that valence and insulator-metal transitions result from 
a change in the occupation numbers of these electronic states, which 
remain themselves basically unchanged in their character. The only 
one quantity that can change in this model is the position
of the $f$-level energy $E_f$. Then, considering a simple
parametrization~\cite{Gon} between the external pressure $p$ and the 
position of $f$-level energy ($E_f \sim p$) all valence 
changes as well as metal-insulator transitions induced by external 
pressure can be directly interpreted in terms of $E_f$. 

The Hamiltonian of the Falicov-Kimball model can be written as the sum of three
terms:

\begin{equation}
H=\sum_{ij}t_{ij}d^+_id_j+U\sum_if^+_if_id^+_id_i+E_f\sum_if^+_if_i,
\end{equation}
where $f^+_i$, $f_i$ are the creation and annihilation
operators  for an electron in  the localized state at
lattice site $i$ with binding energy $E_f$ and $d^+_i$,
$d_i$ are the creation and annihilation operators
of the itinerant spinless electrons in the $d$-band
Wannier state at site $i$.

The first term of (1) is the kinetic energy corresponding to
quantum-mechanical hopping of the itinerant $d$ electrons
between sites $i$ and $j$. These intersite hopping
transitions are described by the matrix  elements $t_{ij}$,
which are $-t$ if $i$ and $j$ are the nearest neighbours and
zero otherwise. The second term represents the on-site
Coulomb interaction between the $d$-band electrons with density
$n_d=N_d/L=\frac{1}{L}\sum_id^+_id_i$ and the localized
$f$ electrons with density $n_f=N_f/L=\frac{1}{L}\sum_if^+_if_i$,
where $L$ is the number of lattice sites. The third  term stands
for the localized $f$ electrons whose sharp energy level is $E_f$.

It is interesting that this simple model is able to explain 
many physical aspects of real mixed valence systems, like the pressure 
induced continuous and discontinuous valence transitions as well 
as discontinuous metal-insulator transitions~\cite{Fark12,aps} despite 
the fact that neglects a direct hybridization (between $d$ and $f$ 
orbitals) which is usually present in these systems. From this point 
of view it is important to ask what happens with the above
mentioned picture of valence and metal-insulator transitions when 
a direct hybridization between the $d$ and $f$ orbitals is 
switched on. For the case of local $d$-$f$ hybridization 
this question has been partially answered in our previous
papers~\cite{Fark3,aps}. 
Using the exact diagonalization and density-matrix-renormalization-group 
calculations we have found that the local hybridization smears
the discontinuous character of valence transitions and stabilizes 
the insulating phase. However,  in real mixed valence materials the
local hybridization is forbidden (for parity reasons) and only the 
nonlocal one (with inversion symmetry) between the  nearest-neighbour
$f$ and $d$ orbitals, is allowed~\cite{Czycholl}. Thus for a correct description
of valence and metal-insulator transitions in these materials one has 
to consider the nonlocal hybridization of the form 
$V_{i,j}=V(\delta_{j,i-1}-\delta_{j,i+1})$ that leads to $k$-dependent 
hybridization of the opposite parity than corresponds to the $d$ band 
($V_k\sim sin(k)$)~\cite{Czycholl}. This type of hybridization has been
used, for example, in our very recent paper~\cite{Fark4} to study the 
formation and condensation of excitonic bound states in the
Falicov-Kimball model and it was shown that the system with 
the nonlocal hybridization exhibits fundamentally different behaviour
than one with the local interaction. While the local hybridization strongly 
supports the formation of excitonc condensate, the nonlocal hybridization
(with inversion symmetry) destroys it completely. 

In the current paper we focus our attention on the problem of 
the pressure induced valence and metal-insulator transitions 
in mixed valence systems with nonlocal hybridization. However,
unlike the previous studies, we use the fundamentally different
approach. Indeed, while in the previous theoretical
works~\cite{Fark12,Fark5} 
the pressure induced valence and metal-insulator transitions  
have been explained via the $E_f-p$ parametrization, in the 
current paper they are interpreted directly via the $V-p$
parametrization. For the case of nonlocal hybridization such 
a parametrization 
is physically very reasonable since the overlap (hybridization) 
of $d$ and $f$ orbitals located on the neighboring lattice sites 
increases with the increasing hydrostatic pressure, due to
the decreasing lattice constant. We show that such a parametrization
leads to a nice qualitative correspondence between the theoretical
results obtained within the proposed approach and experimental
measurements of the pressure dependence of valence and energy
gap in SmB$_6$,  the compound at present very intensely studied 
as a potential candidate of a strongly correlated topological 
insulator~\cite{Nature}. 

\section{Results and Discussion}
In order to investigate  possibilities for valence and insulator-metal 
transitions driven by the nonlocal hybridization (pressure) in the extended
Falicov-Kimball model we have used the density-matrix-renormalization-group
(DMRG) method~\cite{White} that allows 
to treat relatively large clusters ($L\sim100$) and still to keep the high 
accuracy of computations. We typically keep up to 128 states per block, 
although in the numerically
more difficult cases, where the DMRG results converge slower, we keep up
to 500 states. Truncation errors, given by the sum of the density matrix 
eigenvalues of the discarded states, vary from $10^{-8}$ in the worse 
cases to zero in the best cases.

Let us first discuss the picture of metal-insulator transitions obtained 
within the Falicov-Kimball model with the nonlocal hybridization.
The nature of the ground state has been identified
through the behavior of the single particle excitation energy defined as
$\Delta(N)=E_G(N+1)+E_G(N-1)-2E_G(N)$, where $E_G(N)$ is the ground
state energy for $N$ electrons (here we consider the half-filed band case
when the number of electrons $N$ equals to the number of lattice sites $L$). 
Then by a definition~\cite{Kennedy}, the metallic state corresponds to $\Delta(L=\infty)=0$ 
and the insulating one to $\Delta(L=\infty) > 0$. 
The results of our numerical calculations for $\Delta$ as a function of
nonlocal hybridization (pressure) are summarized in Fig.~1 for several
different values of the f-level position $E_f$ and several different
cluster sizes $L$. 
\begin{figure}[h!]
\begin{center}
\includegraphics[width=7cm]{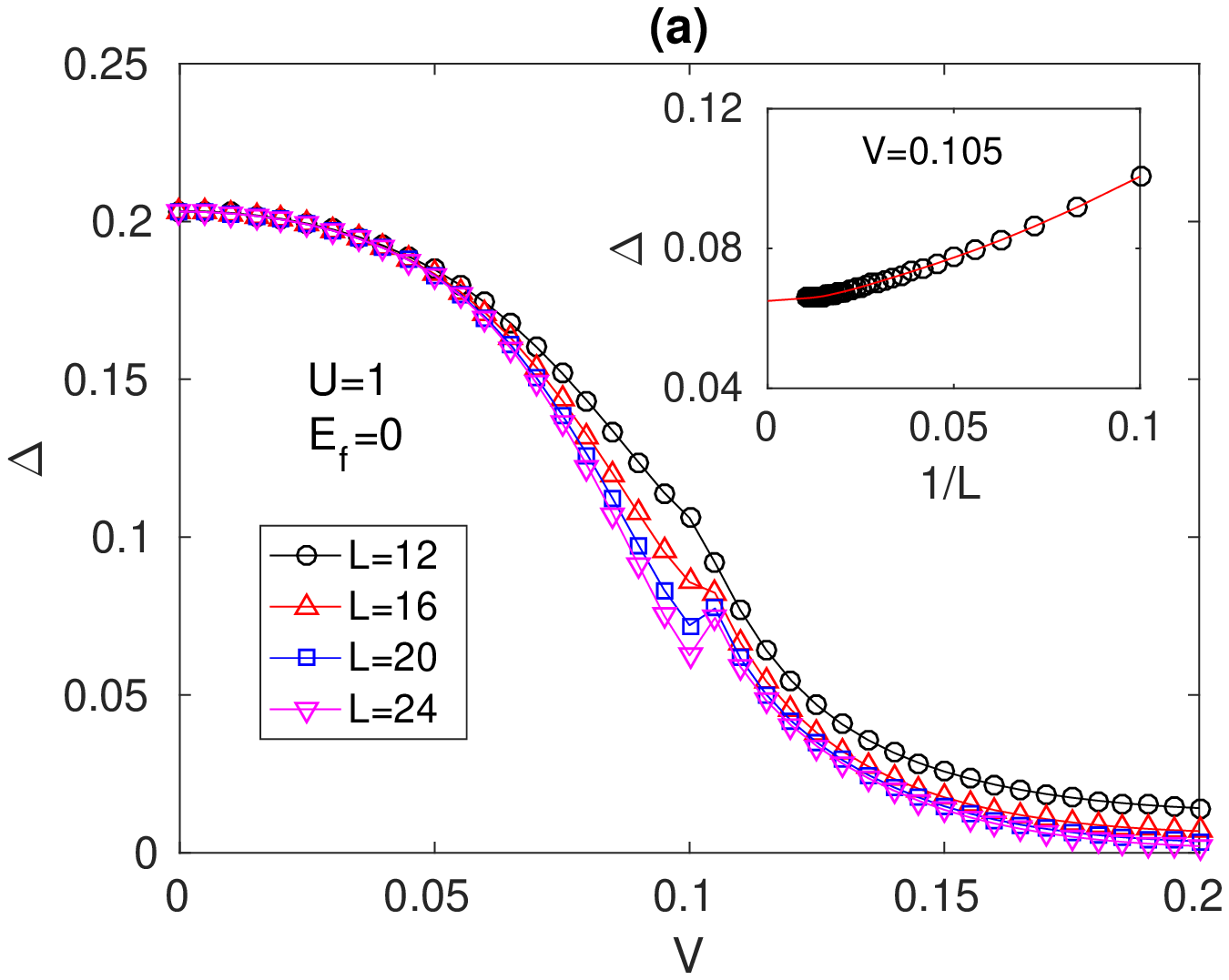}
\includegraphics[width=7cm]{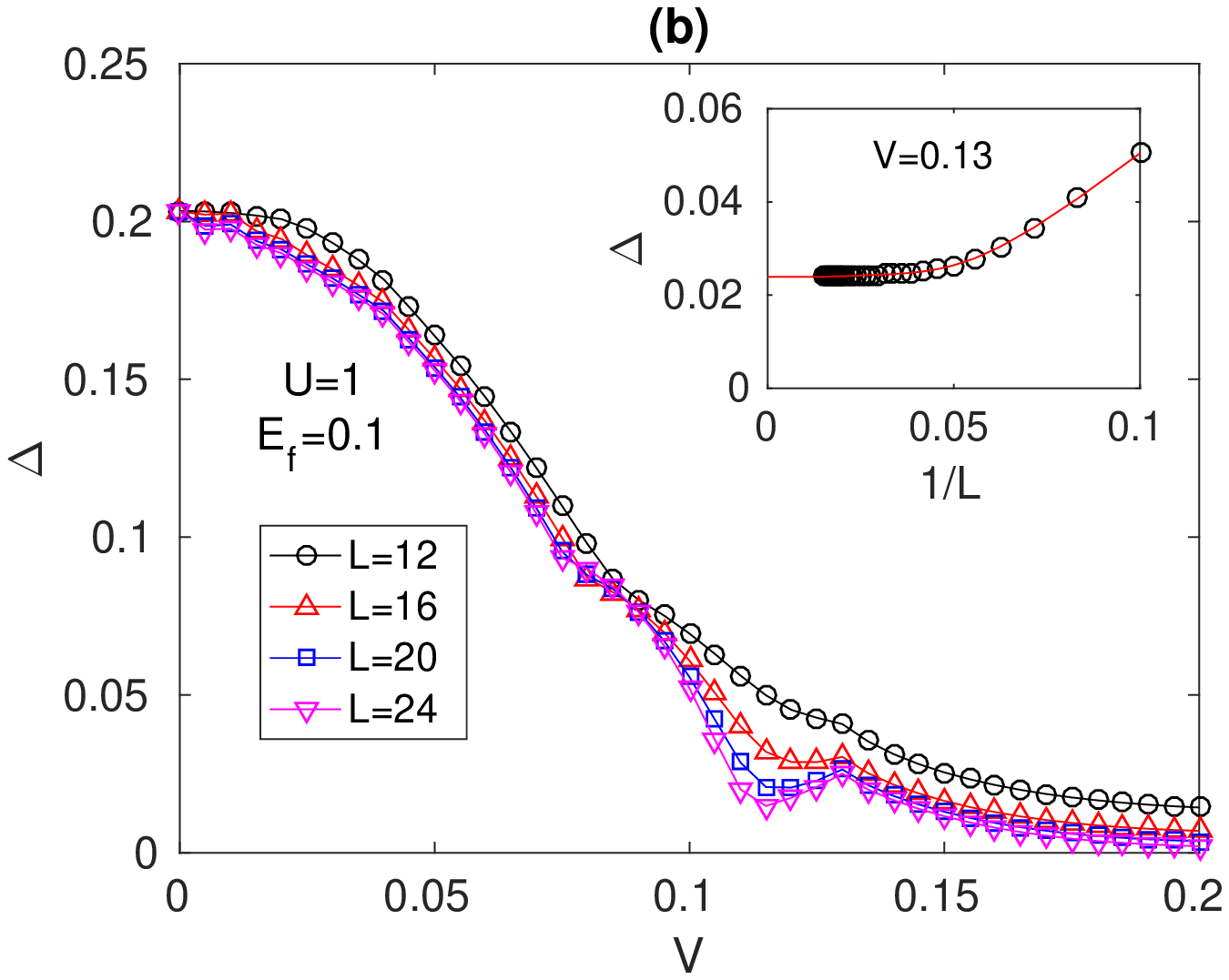}
\includegraphics[width=7cm]{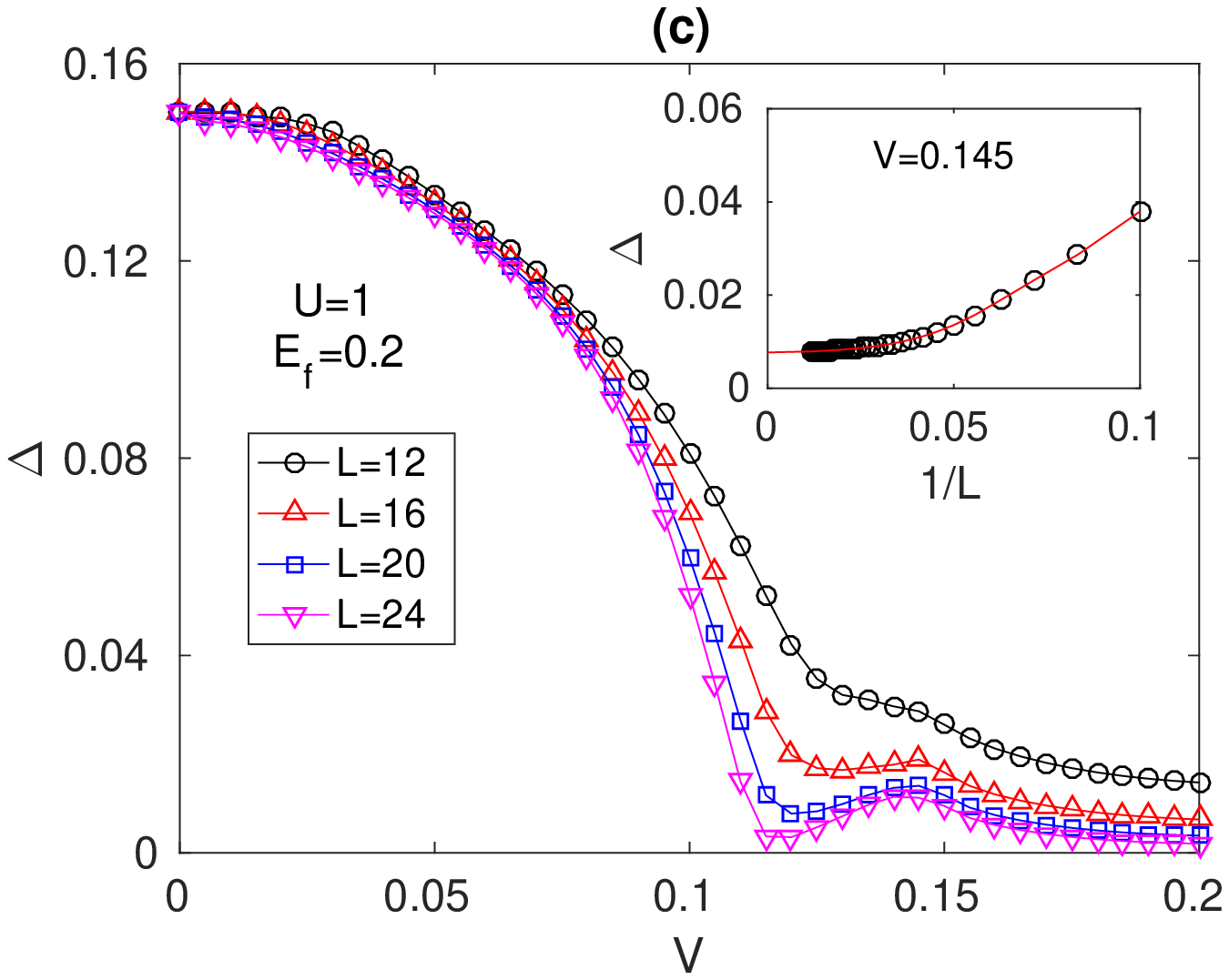}
\includegraphics[width=7cm]{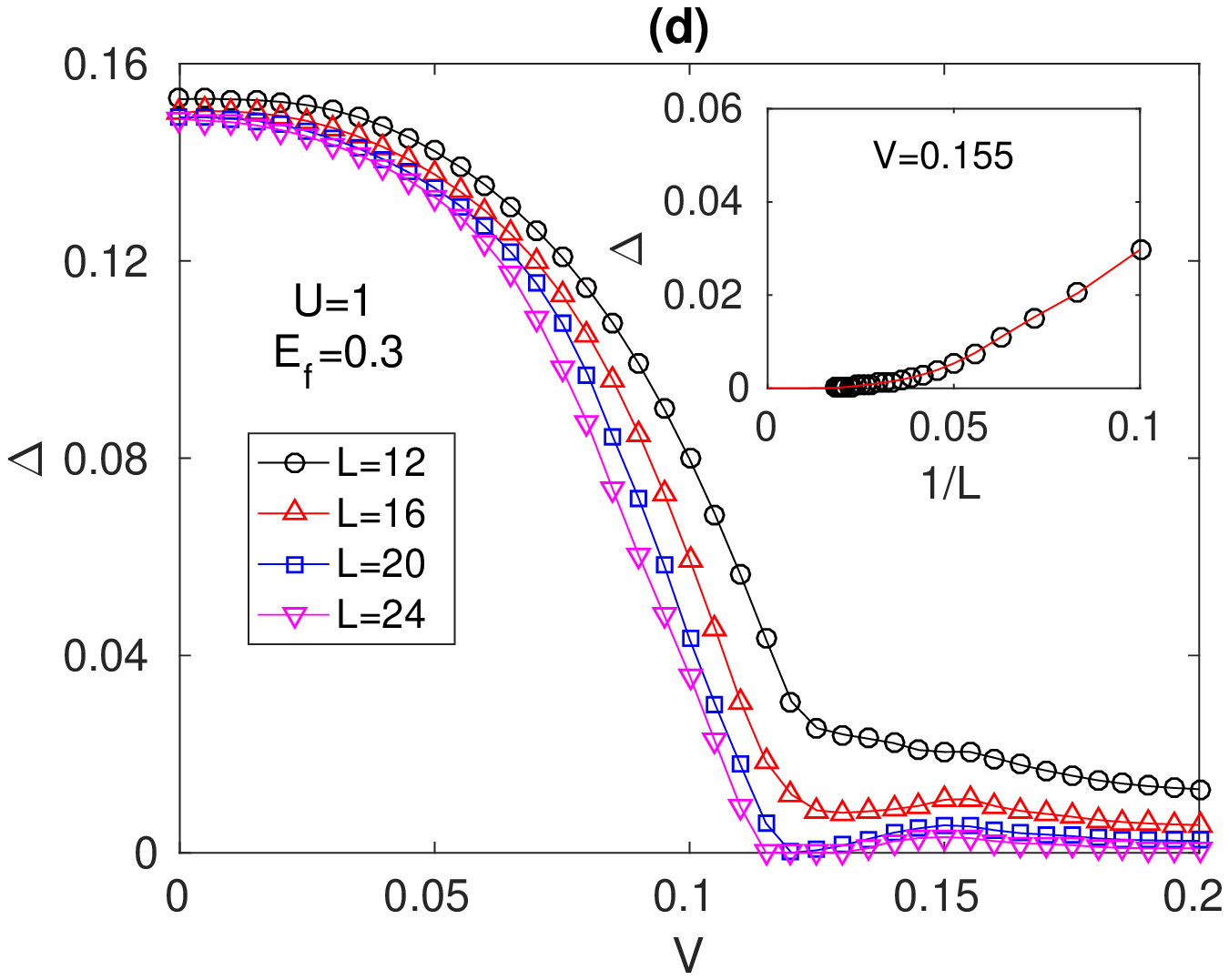}
\end{center}
\vspace*{-0.8cm}
\caption{\small 
The single particle excitation energy $\Delta$ as a function of the
nonlocal hybridization $V$ calculated for several different values of the
$f$-level energy $E_f$ on several different clusters at $U=1$. The insets
show the $1/L$ dependence of $\Delta$ at the position of the local maximum.}
\label{fig3}
\end{figure}
One can see that finite size effects are still
present, but despite this fact the general trends in behaviour
of the single particle excitation energy $\Delta$ as a function of 
nonlocal hybridization $V$ are obvious. The most obvious of them 
is the existence of a critical hybridization $V_c \sim 0.1$,
below and above which the single particle excitation energy
behaves fully differently. Indeed, for $V<V_c$ the single particle 
excitation energy is gradually reduced with increasing $V$, but at 
$V=V_c$ it starts to increase again, reaches its local maximum 
at $V_0$ and then continuously vanishes. The insulating 
character of the ground state below $V_c$ is obvious, however the same 
can not be said for $V>V_c$, where relatively strong finite-size
effects are observed. To reveal the nature of the ground state
in this region, we have performed a detailed finite-size scaling 
analysis of the $L$ dependence of $\Delta$ at the point of
local maximum $V_0$. The resultant behaviors of $\Delta$
as a function of $1/L$ are shown in the insets to Fig.~1 and they
clearly demonstrate that the ground state of the model at $V=V_0$
is insulating for $E_f=0,0.1,0.2$ and metallic for $E_f=0.3$.
This indicates that there exists some critical value of the
$f$-level energy $E_f=E^c_f \sim 0.3$, above which the 
system exhibits by the  nonlocal hybridization induced 
the insulator-metal transition at the critical point $V=V_c$.
To verify this conjecture we have performed (for $E_f=0.3$) 
the exhaustive finite-size scaling analysis of the $L$ 
dependence of $\Delta$ for a wide set of $V$ values from
the region $V<V_c$ as well as $V>V_c$. The results of 
numerical calculations are displayed in Fig.~2a and they
fully confirm our conjecture, at $V=V_c$ the system 
undergoes the insulator-metal transition induced by nonlocal 
hybridization $V$. 
\begin{figure}[h!]
\begin{center}
\includegraphics[width=7cm]{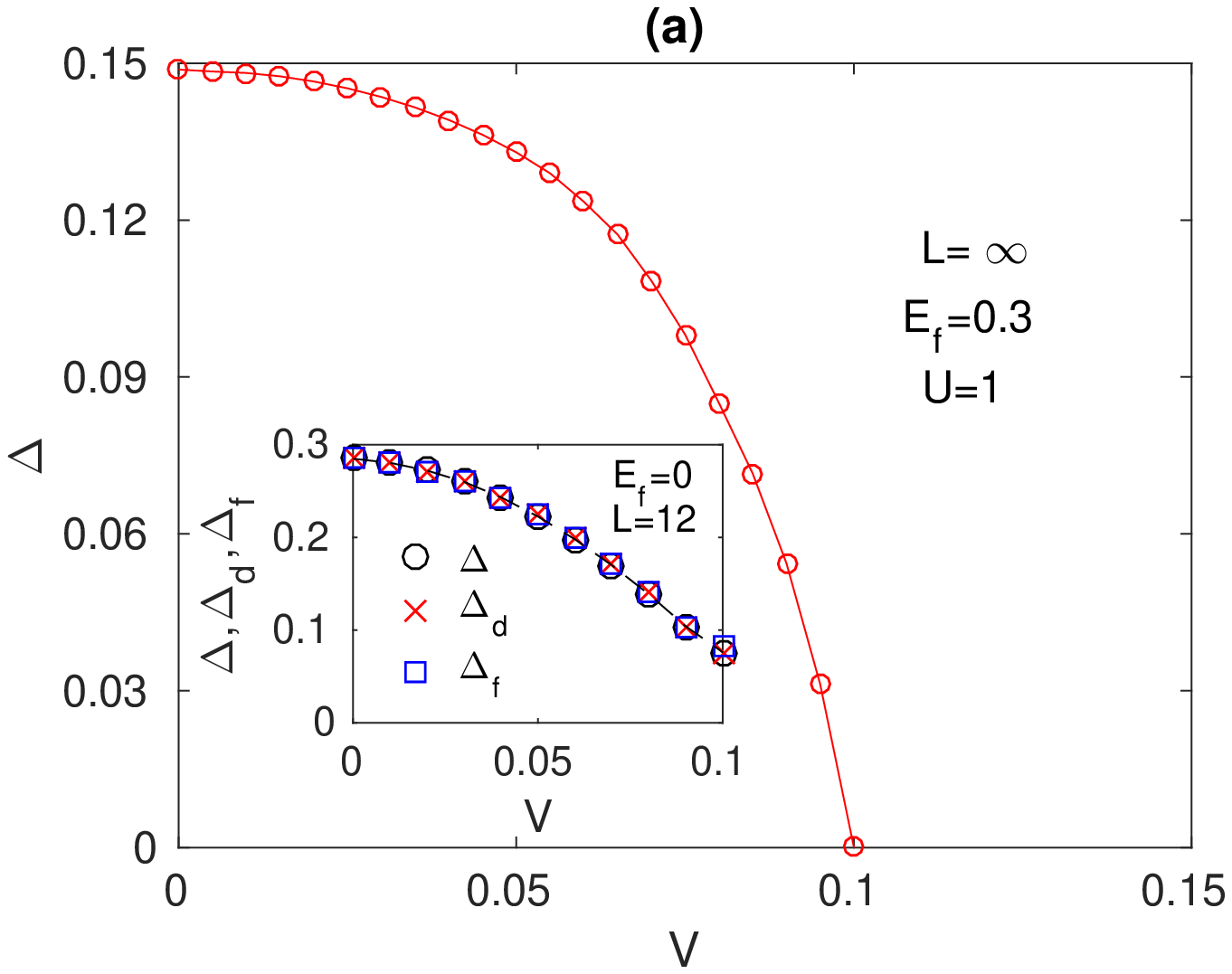}
\includegraphics[width=7cm]{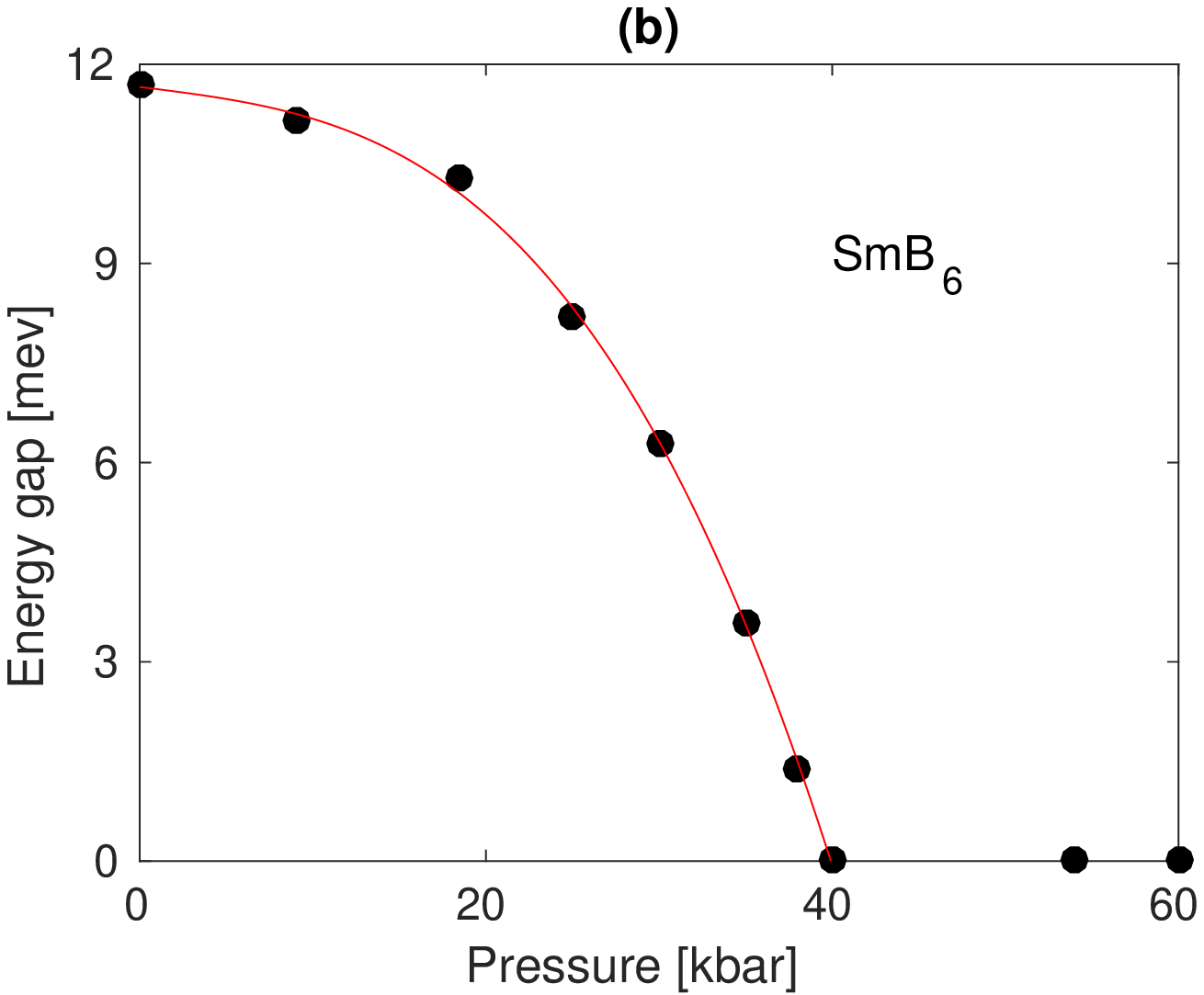}
\end{center}
\vspace*{-0.8cm}
\caption{\small (b) The extrapolated single particle excitation energy $\Delta$ 
as a function of the nonlocal hybridization $V$ calculated for $E_f=0.3$ and $U=1$. 
The inset show $\Delta,\Delta_{f}$ and $\Delta_{d}$ as function of $V$ 
calculated at $E_f=0, U=1$ and $L=12$. (b) Pressure dependence of the 
energy gap in SmB$_6$~\cite{Gabani}.
}
\label{fig3}
\end{figure}
A direct comparison of  our numerical results, 
obtained for the single particle excitation energy $\Delta$ as 
a function of the nonlocal hybridization (see Fig.2a), with the experimental 
ones obtained for the energy gap as a function of pressure
in the SmB$_6$ compound~\cite{Gabani} (see Fig.2b), 
shows that there is a nice correspondence  
between the theoretical and experimental results when one takes
into account the $V-p$ parametrization instead of the $E_f-p$
parametrization. Of course, this comparison will be correct only
if we show that the single particle excitation energy 
is proportional to the energy gap at the Fermi level $E_F$.
For this reason we have calculated directly the single particle
excitation energy $\Delta$ and the $f$ and $d$ electron density 
of states and the corresponding $f$ and $d$ electron gap 
($\Delta_f$ and $\Delta_d$ ) at the Fermi level $E_F$ 
on the small finite cluster of $L=12$ sites. The results
of numerical calculations obtained for several different
values of nonlocal hybridization are shown in the inset 
to Fig.~2a and they clearly demonstrate that $\Delta,\Delta_f,\Delta_d$
coincide for all examined values of nonlocal hybridization,
what validates the above discussed correspondence.

Let us now turn our attention to the problem of valence transitions 
in the Falicov-Kimball model with the nonlocal hybridization in
the context of $V-p$ parametrization. Our DMRG results  
for the $f$-electron occupation number $n_f$ obtained for several
different values of $f$-level position $E_f$ are shown in Fig.~3a.
\begin{figure}[h!]
\begin{center}
\includegraphics[width=7cm]{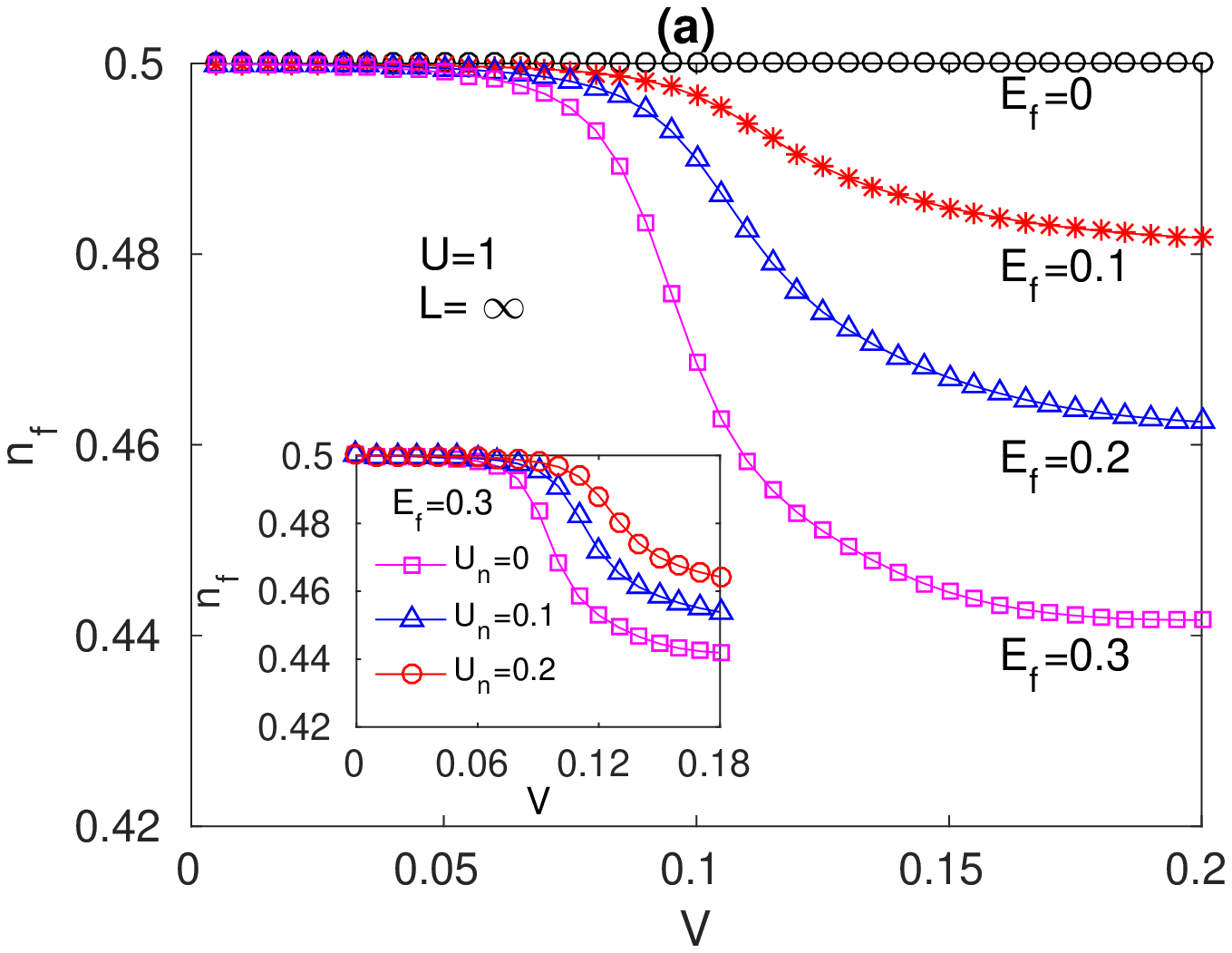}
\includegraphics[width=7cm]{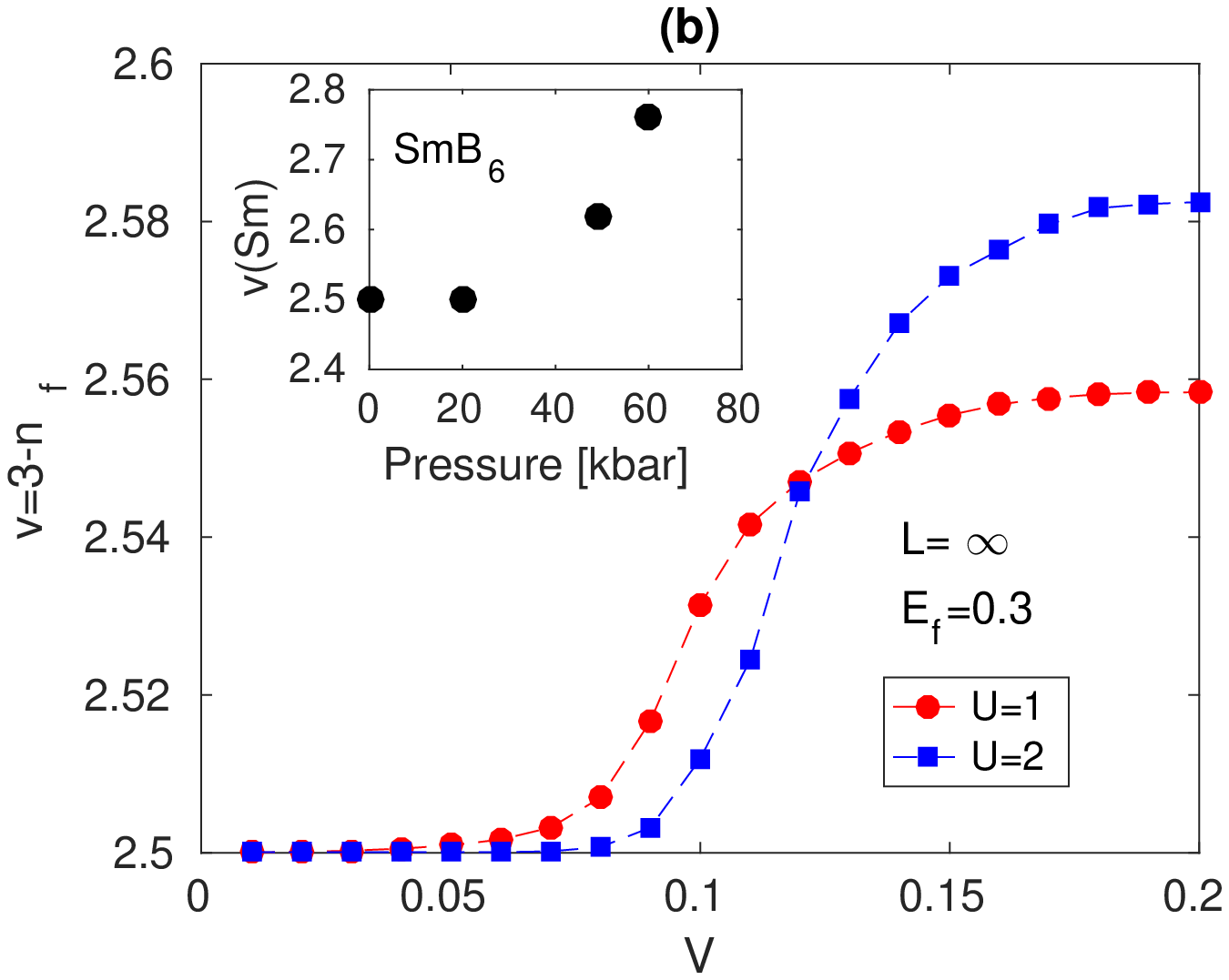}
\end{center}
\vspace*{-0.8cm}
\caption{\small (a) The $f$-electron occupation $n_f$ as a function
of nonlocal hybridization $V$ calculated for several different 
values of the $f$-level energy $E_f$ at $U=1$ and $L=\infty$.
The inset shows the $f$-electron occupation $n_f$ as a function
of $V$ calculated for several different values of the nonlocal 
Coulomb interaction $U_n$ between the $f$ and $d$ electrons 
at $E_f=0.3$, $U=1$ and $L=\infty$.
(b) The valence $v=3-n_f$ as a function of nonlocal hybridization
calculated for two different values of the Coulomb interaction $U$
at $E_f=0.3$ and $L=\infty$. The inset shows the pressure dependence 
of samarium valence in SmB$_6$~\cite{Pristas}. 
}
\label{fig3}
\end{figure}
Comparing these results with ones obtained in our previous paper
within the $E_f-p$ parametrization~\cite{Fark5} one can find the significant
difference. Indeed, while in the case of the $E_f-p$ parametrization
only very small effects of nonlocal hybridization on the 
behavior of the $f$-electron occupation number $n_f$  (valence
transitions) have been observed (especially for $V<V_c$),
the $V-p$ parametrization is able to produces large valence 
changes over the relatively narrow interval of $V$ (p) values.
From  the point of view of above discussed correspondence between 
the theoretical results obtained for the single particle excitation
energy within the $V-p$ parametrization and experimental
results obtained for the pressure dependence of the energy
gap in SmB$_6$, it is interesting to make the same comparison
also for the $f$-electron occupation number $n_f$. Since 
the valence $v$ of the rare-earth ion in the mixed valence systems 
is directly connected with the $f$-electron occupation number $n_f$
(e.g., $v=3-n_f$ for SmB$_6$) such a comparison is possible
thanks experimental data presented in~\cite{Pristas} for the pressure
dependence of samarium valence in SmB$_6$ compound. Both the theoretical
and experimental results for the averaged valence are compared in Fig.~3b
and again one can find a good correspondence between the theoretical 
and experimental results when the $V-p$ parametrization is considered.
To reveal the effects of Coulomb interaction on the valence transitions
we have plotted in Fig.~3b the hybridization dependence of the average
valence $v=3-n_f$ for two different values of the Coulomb interaction
$U=1$ and $U=2$. The results obtained show that the correlation effects
(the Coulomb interaction) enhance considerably the average valence of 
rare-earth ions in mixed valence systems.

Since our model considers only the local Coulomb interaction,
it is natural to ask, if the above described picture of valence
transitions persists also in the physically more realistic case,
when the nonlocal Coulomb interaction between the $d$ and $f$ orbitals
is switched on. To answer this question we have extended our model 
Hamiltonian (1) by the nonlocal nearest-neighbour $d$-$f$ Coulomb 
interaction $H_n=U_n\sum_{i,j}d^+_id_if^+_jf_j$
and calculated the $f$-electron occupation number $n_f$ as a function of
$V$ for several different values of $U_n$. The results of our numerical
calculations are presented in the inset to Fig.~3a and they clearly
demonstrate that results obtained for $U_n=0$ persist also
for finite and physically reasonable values of $U_n$ ($U_n\sim V$).

In summary, the DMRG method is used to examine effects of 
nonlocal hybridization on valence and metal-insulator transitions
in mixed valence systems. On the basis of our numerical results
we have proposed a simple, but very realistic, model for a description 
of pressure induced valence and metal-insulator transitions
in these systems. Instead of the standard $E_f-p$ 
parametrization we have considered the physically more realistic 
$V-p$ parametrization. Under this
assumption the model is able to describe, at least qualitatively, 
the valence as well as  metal-insulator transitions driven by the
external pressure observed experimentally in some rare-earth
systems, like SmB$_6$.

\vspace{0.5cm}
{\small This work was supported by the Slovak Grant Agency VEGA under Grant
2/0112/18.
Calculations were performed in the Computing Centre of the Slovak Academy
of Sciences using the supercomputing infrastructure acquired in
project ITMS 26230120002 and 26210120002 (Slovak infrastructure for
high-performance computing) supported by the Research and Development
Operational Programme funded by the ERDF.}

\end{document}